\documentclass[12pt]{article} 
\usepackage{epsf}
\usepackage{amssymb}
\usepackage{chicago}
\usepackage{fullpage}

\newtheorem{theorem}{Theorem}
\newcommand{\proof}{{\bf Proof:\ \ }}
\newtheorem{exam}{Example}
\newtheorem{prop}{Proposition}
\newtheorem{lemma}{Lemma}
\newtheorem{remark}{Remark}
\newtheorem{observation}{Observation}
\newtheorem{definition}{Definition}

\newcommand{\vol}{{\mbox{\it vol}}}
\newcommand{\old}[1]{}
\newcommand{\rnn}{{{\rm I \! R}^{+}_{0}}}
\newcommand{\qed}{\hfill\hbox{$\quad \Box$ \vspace{1ex}}}
\newcommand{\clique}[1]{\mbox{$E_{#1}$}}
\def\part{\dot\cup}
\def\N{{\rm I\kern-.18em N}}
\def\Q{{\rm {\sf I}\kern-.42em Q}}
\def\Qp{\Q^{\scriptstyle +}}
\def\Qnn{\Qp_{\scriptstyle 0}}

\begin{document}

\title{A Combinatorial Characterization of 
Higher-Dimensional Orthogonal Packing}
 
\author{S\'andor P. Fekete\\
Department of Mathematical Optimization\\
Braunschweig University of Technology\\
D--38106 Braunschweig\\
GERMANY \\
{\tt s.fekete@tu-bs.de}\\
\and
        J\"org Schepers\\
        IBM Germany\\
        Gustav-Heinemann-Ufer 120/122\\
        D--50968 K\"oln\\
        GERMANY\\
        {\tt schepers@de.ibm.com}
}

\date{}
\maketitle 

\begin{abstract}
Higher-dimensional orthogonal packing problems have a wide range of 
practical applications, including packing, cutting, and scheduling.
Previous efforts for exact algorithms have been unable
to avoid structural problems that appear for instances
in two- or higher-dimensional space.
We present a new approach for modeling packings,
using a graph-theoretical characterization of feasible packings.
Our characterization allows it to deal with classes of packings
that share a certain combinatorial structure,
instead of having to consider one packing at a time. In addition,
we can make use of elegant algorithmic properties
of certain classes of graphs.
This allows our characterization to be the basis for a 
successful branch-and-bound framework.

This is the first in a series of papers describing
new approaches to higher-dimensional packing.
\end{abstract}



\section{Introduction}
\label{sec:intro}
The problem of cutting a rectangle into smaller rectangular 
pieces of given sizes is known as 
the {\em two--dimensional cutting stock problem}. It arises in many 
industries, where 
steel, glass, wood, or textile materials are cut, but it also occurs 
in less obvious contexts, such as
machine scheduling or optimizing the layout of advertisements in newspapers.
The three-dimensional problem is important for practical 
applications such as container 
loading or scheduling with partitionable resources. It can be 
thought of as packing boxes into a container. 
When arranging these boxes in a container $C$, we have to 
preserve the orientations of the boxes; this constraint usually arises from 
considerations for stability in packings (``this side up''), 
from asymmetric
texture of material in cutting-stock problems, or from different types
of coordinates in scheduling problems. In the context of technical computer 
science, \citeN{Teich} %xxx
consider an application of three-dimensional packing 
to dynamic reconfiguration of hardware; 
one of the axes corresponds to time, the two others
to different spatial dimensions.

We refer to the generalized problem in $d \ge 2$ dimensions as 
the {\em $d$-dimensional orthogonal knapsack problem (OKP-$d$)}.
Being a generalization of the bin packing problem (BPP),
the OKP-$d$ is ${\cal NP}$-complete in the strict sense -- see
\citeN{GAJO79}. 
%(Note that we consider the constrained problem, where bounds on the number of 
%rectangles of each size are imposed.)
The vast majority of work done in this field refers to a restricted problem, 
where only so--called  {\em guillotine patterns} are permitted. 
This constraint arises from certain industrial cutting applications: 
guillotine patterns are those packings that can be generated by
applying a sequence of edge-to-edge cuts;
see \citeN{CHWH77} and \citeN{WANG83}.
The recursive structure of these patterns makes
this variant much easier to solve than {\em general}
 or {\em non--guillotine} problems. 

Relatively few authors have dealt with the exact 
solution of non--guillotine problems.
All of them focus on the problem in two dimensions.
%Bir\'{o} and Boros (1984) 
\citeN{BIBO84} gave a 
characterization of non--guillotine patterns using
network flows but derived no algorithm. %Dowsland (1987) 
\citeN{DOWS87} 
proposed an exact algorithm for the case that all boxes have equal size. 
%Arenales and Morabito (1995) 
\citeN{ARMO95} extended an approach
for guillotine problems to cover a certain type of non--guillotine patterns. 
So far, only two exact algorithms have been proposed 
and tested for the general case of knapsack problems. 
%Beasley (1985) 
\citeN{BEA85}
and 
%Hadjiconstantinou and Christofides (1995) 
\citeN{HACHR95} have 
given different 0--1 integer programming formulations for this problem. 
Even for small problem instances, 
they have to consider very large 0--1 programs,
because the number of variables depends on the size of the container that is to be
packed. The largest instance that was solved in either article has 9 out of 22 boxes packed into
a 30 $\times$ 30 container. (See the ``OR-Library'' by
%Beasley 
\citeN{ORLIB90} for a number of test instances.)
After an initial reduction phase,
\citeN{BEA85} gets a 0-1 program with over 8000 variables and more than
800 constraints; the program by   
Hadjiconstantinou and Christofides still contains more than 
1400 0--1 variables and over 5000 constraints. 
{From} Lagrangian relaxations, they derive upper bounds for a 
branch--and--bound algorithm that are
improved using subgradient optimization.
The process of traversing the search tree corresponds to the iterative 
generation of an optimal packing. 

We should note that more recently, two papers on the related problem
of two- and three-dimensional bin packing have been presented:
%Martello and Vigo (1996) 
\citeN{MAVI96}
consider the two-dimensional case,
while 
%Martello, Pisinger, and Vigo (1997) 
\citeN{MAPV97} deal
with three-dimensional bin packing. We discuss aspects
of those papers in %our second paper~
\cite{pack2}, when considering
bounds for higher-dimensional packing problems.
Even more recently, %Padberg~
\citeN{Pad00} gave a mixed integer
programming formulation for three-dimensional packing problems,
similar to the one anticipated by the second author
in his thesis
\cite{Sch97}.
Padberg expressed the hope that using a number of techniques from
branch-and-cut will be useful; however, he did not provide
any practical results to support this hope.

In this paper we describe a different approach to characterizing
feasible packings and constructing optimal solutions.
We use a graph-theoretic characterization of the relative position
of the boxes in a feasible packing. This allows a much more efficient
way to construct an optimal solution for a problem instance:
combined with good heuristics for dismissing infeasible subsets
of boxes, we develop a two-level tree search. This exact algorithm
has been implemented; it outperforms previous methods by a wide margin.

The rest of this paper is organized as follows: after a formal problem
description in Section~\ref{sec:prelim},
we introduce the novel concept of packing classes  in Section~\ref{se:pack}
and show that it suffices to deal with the existence of feasible packings.
In Section~\ref{sec:algo}, we describe algorithmically useful properties
of packing classes, and highlight algorithmic difficulties in Section~\ref{se:npc}.
In \cite{pack3}, we give a detailed description
of how packing classes can be combined with other algorithmic ideas
for obtaining powerful exact algorithms for orthogonal packing problems.

\section{Preliminaries}
\label{sec:prelim}
Why are two- and higher-dimensional packing problems harder to solve
than their one-dimensional counterparts?

In order to decide whether another item can be packed into
a partially filled container, in the one-dimensional case,
we only need to know the total size of the items that have
already been placed.
When dealing with a two-
or higher-dimensional problem, we also need to know the arrangement
of boxes, i.e., their {\em packing}.
Thus, the performance of an algorithm for higher-dimensional
packing hinges on the use of an appropriate representation
of packings.

A main difficulty is highlighted by the following easy observation:
In general, the feasible space for packing further objects into a
container is not convex. This means that we cannot simply
formulate a higher-dimensional orthogonal packing problem
as a straightforward integer linear program.
Nevertheless, %Beasley~
\citeN{BEA85} and 
%Hadjiconstantinou and Christofides~
\citeN{HACHR95} have used
such a formulation, where the free space is controlled
with the help of a large number of 0,1-variables, describing
a partition of the feasible space into convex pieces.
They only consider the two-dimensional case.
These formulations lead to very large
0,1-programs, even for packing problems of moderate size.
It seems hopeless to try this approach for three-dimensional 
instances of relevant size.

We present an alternative approach to these 0,1-formulations.
Our new way of modeling packings is based on a graph-theoretic
characterization of packings. It can be used for any number of 
dimensions and leads to the central term of {\em packing class},
which describes a whole family of combinatorially equivalent feasible
packings. 
The implementation of the resulting algorithms
is described in detail in \cite{pack3}.

We start by introducing some basic notation and definitions.

\subsection{Problem Description and Notation}
%\subsection*{Eingabedaten}
In the following, we consider a set of items 
that need to be packed into a {\em container}. 
We concentrate on $d$-dimensional boxes, and the items
are also called {\em boxes}.

The input data for a $d$-dimensional orthogonal packing problem
is a finite set of boxes $V$, and a (vector-valued) {\em size function}
$w: V \rightarrow {\Qnn}^{d}$ that describes the size
of each box in any dimension $x_1,\ldots, x_n$.
For the orthogonal knapsack problem (OKP), we also have 
a value function $v: V \rightarrow {\Qp}$ that describes
the objective function value for each box.

%By grouping together boxes with identical input data, i.\,e., boxes
%with identical size and value, we get a partitioning 
%$$
%V = \part_{t=1}^{m} T_{t}.
%$$
%of the set of boxes. These sets 
%$T_{t}$ are called {\em box types}.

The size of the container is given by a vector
$W \in {\Q^{+}}^{d}$. 
%Whenever convenient, we may assume that
%the container is a $d$-dimensional unit cube.
Without loss of generality, we assume that each individual box fits
into the container, i.\,e., $ w(b) \le W $ holds for each box. 

For the volumes of boxes and container we use the following notation.
If $b \in V$, and $w$ is a size function defined on $V$, then
$$
\vol_w(b) := \prod_{i=1}^{d} w_{i}(b)
$$
denotes the volume of box $b$ with respect to $w$. Similarly, the
volume of the container is denoted by
$$
\vol_W := \prod_{i=1}^{d} W_{i}.
$$

If $S$ is a finite set and $f$ a real-valued function on $S$,
then we use the abbreviation
$$
f(S) := \sum_{x \in S} f(x).
$$

Finally, we use the notation $\{v,w\}$ for an edge in an undirected graph;
we write $(v,w)$ when referring to a directed edge.
For a given set of vertices $S\subseteq V$ in a graph
$G=(V,E)$, an {\em induced subgraph} is given by
$G[S]=(S,E[S])$, with $E[S]$ being the set of edges
in $G$ that connect vertices in $S$. For
a vertex $v\in V$ in a directed graph $G=(V,E)$,
$\delta^-(v)$ denotes the indegree of $v$.
More graph-theoretic
terms will be introduced when needed. In all cases, see the book by
%Golumbic~
\citeN{GOL80} for more background.

\subsection{Orthogonal Packings}
We consider arrangements of boxes that satisfy the following constraints:
\begin{enumerate}
\item {\bf Orthogonality:} Each face of a box is parallel to
a face of the container.
\item {\bf Closedness:} No box may exceed the boundaries of the container.
\item {\bf Disjointness:} No two boxes may overlap. 
\item {\bf Fixed Orientations:} The boxes must not be rotated.
\end{enumerate}
In the following, we imply these conditions when ``packing boxes into 
a container'', ``considering a set of boxes that fits into a
container'', and speak of {\em packings}.

In order to formalize the notion of a packing, we introduce a
Cartesian coordinate system with axes parallel to the edges of the
container. The origin is one corner of the container, which is fully
contained in the positive orthant. Thus, a packing is a function
that maps each box to the coordinates of the one of its corners
that is closest to the origin.
\begin{definition}[Packing] $\,$\\
Given a set of boxes $V$, a size function $w$, and a container size $W$.

A function
$p: V \rightarrow {\Qnn}^{d}$ is called a {\em packing} of $(V,w,W)$, iff
\begin{eqnarray}  
\forall b \in V : && p(b)+ w(b) \le W \label{defpm1} \\
\forall b,c \in V, b \ne c,  \exists i \in \{1,\dots,d\}: &&
I^{p}_{i}(b) \cap
I^{p}_{i}(c) 
  = \emptyset .
\label{defpm2}  
\end{eqnarray}  
\end{definition}
Here, $I^{p}_{i}(b)$
denotes the interval $[\, p_{i}(b), p_{i}(b) + w_{i}(b)  \, )$,
for a box $b \in V$, and a direction $i \in \{1,\dots,d\}$.

Condition (\ref{defpm1}) implies closedness,
while (\ref{defpm2}) implies disjointness of the boxes.
Because components of the size function are fixed, we pack with
fixed orientations.

\begin{remark}
The condition ``p is a packing for $(V,w,W)$'' can be expressed
by linear constraints. For this purpose,
condition (\ref{defpm2}) has to be replaced by
$$
\forall b,c \in V, b \ne c  \bigvee_{i=1}^{d} \,
p_{i}(b) + w_{i}(b) \le  p_{i}(c) \, \vee \,
p_{i}(c) + w_{i}(c) \le  p_{i}(b). 
$$
\end{remark}
In  \citeN{NEWO88}, p.~12f., it is described how the disjunction
can be transformed into $2d+1$ linear inequalities by
using $2d$ additional 0,1-variables. The resulting
mixed integer programs seem to be beyond practical size.
Our own experience with a mixed integer approach has been
far from promising, see~\citeN{Sch97}; 
it remains to be seen whether the more recent revival by
\citeN{Pad00} will lead to any computational results.

\subsection{Objective Functions}
Depending on the objective function, we distinguish 
three types of orthogonal packing problems:

\begin{definition}[Optimization Problems]\ \\
\label{def:obj}
\vspace*{-5mm}
\begin{itemize}
\item The {\bf Strip Packing Problem (SPP)} asks for the minimal height
$W_d$ of a container that can hold all boxes, where the size in the other
$d-1$ dimensions $W_1,\ldots, W_{d-1}$ are fixed.
\item 
For the 
{\bf Orthogonal Bin Packing Problem (OBPP)}, we have to determine
the minimal number of identical containers that are required
to pack all the boxes.
\item In the {\bf Orthogonal Knapsack Problem (OKP)},
each box has an objective value. A container has to be packed
such that the total value of the packed boxes is maximized.
\end{itemize}
\end{definition}

This classification follows the scheme introduced by
\citeN{WOT96}. (He introduces a fourth class called 
{\bf Pallet Loading}, where all boxes have identical size and value $1$;
this is a special case of the knapsack problem.)
To clarify the dimension of a problem, we may speak of
OKP-$2$, OKP-$3$, OKP-$d$, etc.

Problems SPP and OBPP are closely related.
For $d \in \N$, an OBPP-$d$ instance can be transformed into a special
type of SPP-$(d+1)$ instance, by assigning the same $x_{d+1}$-size
to all boxes. This is of some importance for deriving relaxations
and lower bounds.

For all orthogonal packing problems we have to satisfy the constraint
that a given set of boxes fits into the container. This subproblem is
of crucial importance for our approach.
\begin{definition}[Orthogonal Packing Problem\ \ (OPP-$d$)]\ \\
For a set of boxes $V$, a size function $w$, and the container size $W$,
is there a packing for $(V,w,W)$?
\end{definition}
Clearly, the OPP is the decision problem for the above optimization problems.

\section{Problem formulation and mathematical approach}
\label{se:pack}

\subsection{Modeling the Problem}\label{model}
In the following, we describe a new way of 
modeling feasible packings. The basic idea is to use the
combinatorial information induced by relative box positions.
For this purpose, we consider projections along the different coordinate
axes; overlap in these projections defines an interval graph for
each coordinate. Using properties of interval graphs and
additional conditions, we can tackle two fundamental
problems of exact enumeration algorithms:

\begin{enumerate}
\item How can we prove in reasonable time that a particular subset of boxes is
infeasible for packing?
\vspace*{1mm}
\item How do we avoid treating equivalent cases more than once?
\end{enumerate}

\vspace*{1mm}
\subsection{Packing classes}\label{oppppc}
\vspace*{1mm}
Instead of dealing with single packings we
handle classes of packings that share a certain combinatorial structure.
This structure arises from the way different boxes in a packing can 
``see'' each other
orthogonal to one of the coordinate axes. (So-called {\em box visibility graphs}
have been considered as means for representing graphs, e.g., see
the references in~\citeN{FEME97}.)
We will see that transitive orientations of certain classes of graphs
correspond to possible packings if and only if specific additional conditions
are satisfied. This allows us to make use of a number of powerful theorems
\cite{Gh62,GiHo64,GOL80,KOMO89} to perform pruning of our 
branch-and-bound tree.

\begin{figure}[hbtp]
\begin{center}
\leavevmode
\epsfxsize=.55\textwidth
\epsffile{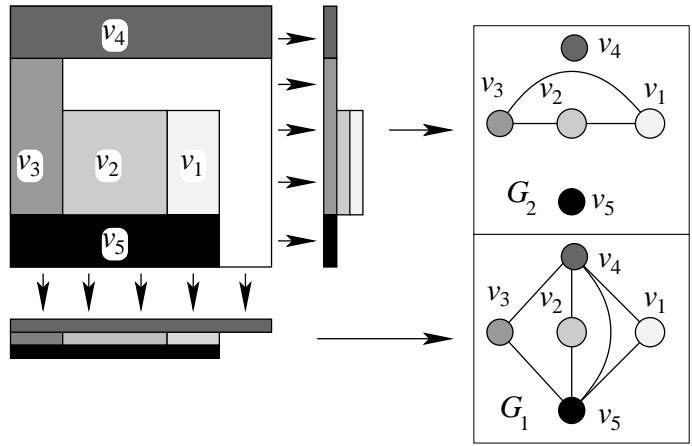}
\caption{A two-dimensional packing and the interval graphs $G_1$, $G_2$ induced by the axis-parallel projections.}
\label{classes}
\end{center}
\end{figure}

\vspace*{1mm}
For a $d$-dimensional packing, consider the projections of the boxes
onto the $d$ coordinate axes $x_i$. Each of these projections induces
a graph $G_i=(V,E_i)$: 
\[\{b,c\}\in E_i \Leftrightarrow I_i^p(b)\cap I_i^p(c)\neq \emptyset.\]
In other words, two boxes are adjacent in $G_i$, if and only if
their $x_i$ projections overlap. (See Figure~\ref{classes} for a 
two-dimensional example.)
By definition, the $G_i$ are interval graphs, thus they have
algorithmically useful properties that are described
in Section~\ref{sec:algo}. %(See~\citeN{GOL80, KOMO89}.)

\vspace*{1mm}
A set of boxes 
$S \subseteq V$ is called 
{\em $x_i$--feasible}, if
$\sum_{b\in S}w_i(b)\leq W_i$,
i.e., if the boxes in $S$
can be lined up along the $x_i$--axis
without exceeding the $x_i$--width of the container.
Finally, it follows from the disjointness condition (\ref{defpm2})
that no two boxes may overlap in all coordinate projections.
Thus, we get the following necessary conditions: 

\begin{observation} 
\label{ob:nec}
For any feasible packing, the graphs $G_i=(V, E_i), i=1,\ldots,d$ 
have the following properties:
\begin{eqnarray*} 
P1:&& \mbox{Each $G_{i}:=(V,E_{i})$ is an interval graph.}
\\
P2:&& \mbox{Each stable set } S \mbox{ of } G_{i} \mbox{ is $x_i$--feasible}. 
\\
P3:&& \bigcap_{i=1}^{d} E_{i} = \emptyset.
\end{eqnarray*}
\end{observation}

The main result of this paper is the converse: These three conditions
are also sufficient.

A family of $E=(E_1,\ldots,E_d)$ of edges sets
for a vertex set $V$ representing a set of boxes
is called a {\em packing class} for $(V,w)$, 
if and only if it satisfies the conditions $P1$, $P2$, $P3$.
For any given $G_i=(V,E_i)$, we denote by $\overline{G_i}$ the complement graph
$(V,\overline{E_i})$ for $G_i$. See Figure~\ref{dings}.
If $G_i$ arises from a packing, any edge in $\overline{G_i}$
corresponds to two boxes with non-overlapping $x_i$-projection,
hence we can think of $\overline{G_i}$ as the comparability graph of the boxes
in direction $x_i$: Two boxes are comparable (and hence there
is an edge in $\overline{G_i}$), if their $x_i$-projections do not
overlap, and one projection lies strictly to the left of the other
projection. 

Next we consider directed graphs arising from some of these undirected 
graphs, indicating the partial order of the $x_i$-projections. 
A directed graph $D=(V,A)$ is called a {\em transitive orientation}
of a graph $G=(V,E)$, iff there is a bijection ${\cal O}:E\rightarrow A$
with the property 
\[(a,b)\in A\wedge (b,c)\in A \Rightarrow (a,c)\in A.\]
Note that this implies that the undirected edge $\{a,c\}$ must
also be contained in $E$. For more background on transitive orientations,
see the book by \citeN{GOL80}.

For any packing class $E$, we consider
transitive orientations $F_{i}$ of the 
comparability graph $\overline{G_{i}}$,
and call $F = (F_{1},\dots,F_{d})$ a {\em transitive orientation}
of the packing class $E$. In the following, we show that
for any transitive orientation $F$ of a packing class, there
is a packing. 
For this purpose, define a mapping $p_{i}^{F}: V \rightarrow \rnn^d$ by
\begin{equation}
p^{F}_{i}(v)  := \max\{p^{F}_{i}(u) + w_{i}(u) | (u,v) \in F_{i} \}
\label{konstr}
\end{equation}
for $v \in V, \, i \in \{1,\dots,d\}$.

\subsection{Sufficiency of Packing Classes}\label{subse:suff}
Without loss of generality, it is sufficient to consider packings 
that are generalized ``bottom-left'' justified, such that
any lower bounding coordinate of a box is at zero
or at the upper bounding coordinate of another box (without requiring
the two boxes to touch). More formally, we say that
a packing is {\em gapless}, if 
\[\quad \forall i \in \{1,\dots,d\} \  \forall v \in V : \quad
p_{i}(v) = 0 \quad \vee \quad
\exists u \in V: \, p_{i}(v) = p_{i}(u) + w_{i}(u).\]
In the following, we show that the above three conditions
on the graphs forming a packing class are not only necessary
for the existence of a packing,
but they are also sufficient, even for the subclass of 
gapless packings. This implies that a packing class is not just
a set of graphs, but can also be interpreted as a whole equivalence
class of feasible packings.

\begin{lemma}\label{pmsatz3}
Any packing class $E$ represents a feasible packing.
More precisely, for a transitive orientation $F$ of $E$,
the function $p^{F}: V \rightarrow \rnn^d$ 
describes a gapless packing.
\end{lemma}

\proof
Consider a packing class $E$. As each $G_i=(V,E_i)$
is an interval graph, the complement $\overline{G_i}$
is a comparability graph, hence it has a transitive
orientation $F_i$. Let $F=(F_1,\ldots,F_d)$ be any such set
of transitive orientations. As defined above, this induces
a mapping $p^F$.
We place all boxes $v$ at the positions given by $p^F(v)$.
To show that $p^F$ is a packing, we have to show that
\begin{enumerate}
\item All boxes are within the boundaries of the container.
\item No two boxes overlap.
\end{enumerate}

To establish 1., we show that at any stage there must be a set of boxes
that determines the overall $x_i$-width of the partial packing, so by
$P2$, the partial packing must fit in all directions.
For this purpose, let $v\in V, i\in\{1,\ldots, d\}$. By construction of $p^F$,
the acyclic digraph $(V,F_i)$ contains a directed
path $(v^{(0)},\ldots,v^{(r)})$ with $\delta^-(v^{(0)})=0$
and $v^{(r)}=v$, such that
\[p^{F}_{i}(v) = \sum_{k=0}^{r-1} w_{i}(v^{(k)}).\]

Because $F_i$ is transitive, $S:=\{v^{(0)}, \ldots, v^{(r)}\}$ induces
a clique in $\overline{G_i}$, implying that $S$ is a stable set in $G_i$.
Hence by condition $P2$,
\[p_{i}^{F}(v) + w_{i}(v) = \sum_{k=0}^{r} w_{i}(v^{(k)}) \le W_i,
\mbox{  implying 1.}\]

To see 2., consider $u,v\in V$, $u\neq v$. By $P3$, there is
an $i\in \{1,\dots,d\}$ with the (undirected) edge
$\{u,v\} \notin E_{i}$, hence
$\{u,v\} \in \overline{E_{i}}$. Because $F_i$ is an orientation of 
$\overline{E_i}$, either $(u,v) \in F_{i}$ or $(v,u) \in F_{i}$.
By construction of $p^F$, we get
\[p^{F}_{i}(v) \ge p^{F}_{i}(u) + w_{i}(u) \quad \lor \quad p^{F}_{i}(u) 
\ge p^{F} _{i}(v) + w_{i}(v).\]
In both cases, $u$ and $v$ can be separated by an $x_i$-orthogonal
hyperplane. Hence, the boxes $u$ and $v$ are disjoint and $p^F$
describes a packing. By construction of $p^F$, the packing is gapless.
\qed

\bigskip
{From} Observation~\ref{ob:nec} and Lemma~\ref{pmsatz3}, the main theorem
of this paper follows:

\begin{theorem} \label{th:main}
A set of $d$-dimensional boxes $V$ can be packed into a container, iff there is
a packing class $E$ for $(V,w)$, i.e., a set of $d$ graphs $G_1,\ldots,G_d$
with the properties P1, P2, P3.
\end{theorem}
 
The proof of Lemma~\ref{pmsatz3} 
shows constructively that {\em any} orientation of a packing class 
corresponds to a packing; basically, an orientation $(v_1,v_2)$
of an edge $\{v_1, v_2\}\in E_i$ means that $v_1$ can ``see'' $v_2$ along
the positive $x_i$-axis. To illustrate the structure, we give
the following example:

\begin{exam}
Consider the OPP-$2$ $P = (V,w)$, defined by
\begin{eqnarray*}
&& V := \{ b_{1},b_{2},b_{3},b_{4},b_{5} \} \\
&& w(b_{1}) :=\left(4,1\right), \quad w(b_{2}):=\left(5,1\right), \quad w(b_{3}):=\left(1,3\right), \\
&& w(b_{4}):=\left(2,2\right), \quad w(b_{5}):=\left(1,2\right), \quad W:=\left(5,5\right).
\end{eqnarray*}
and the packing class $E$ for $P$ as shown in Figure~\ref{classes}.
(Note that box $i$ is darker than box $j$ whenever $i<j$.)

\begin{figure}[htbp]
\begin{center}
\leavevmode
\epsfxsize=.6\textwidth
\epsffile{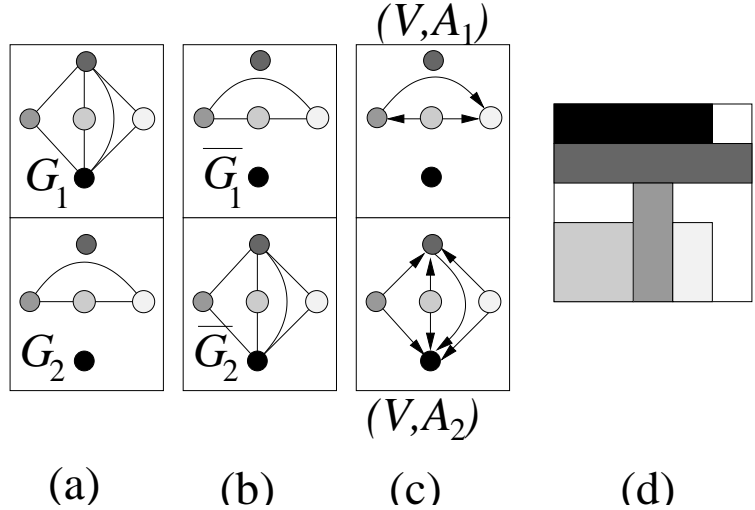}
\caption{\label{dings} Constructing a packing for a two-dimensional
packing class $E$: (a) A pair of interval graphs. 
(b) The corresponding complement, i.e., comparability graphs. 
(c) Transitive orientations of
the comparability graphs. (d) A packing built from the
transitive orientations. It is feasible by properties $P2$ and $P3$.}
\label{fi:orient}
\end{center}
\end{figure}

The complements of the component graphs $E_{1}$ and $E_{2}$
from the example in Figure~\ref{dings} each have six transitive orientations.
Hence there are 36 transitive orientations of $E$.
Figure~\ref{pmpic2} shows the corresponding packings,
constructed by virtue of (\ref{konstr}).

\begin{figure}[htbp]
\begin{center}
\leavevmode
\epsfxsize=.55\textwidth
\epsffile{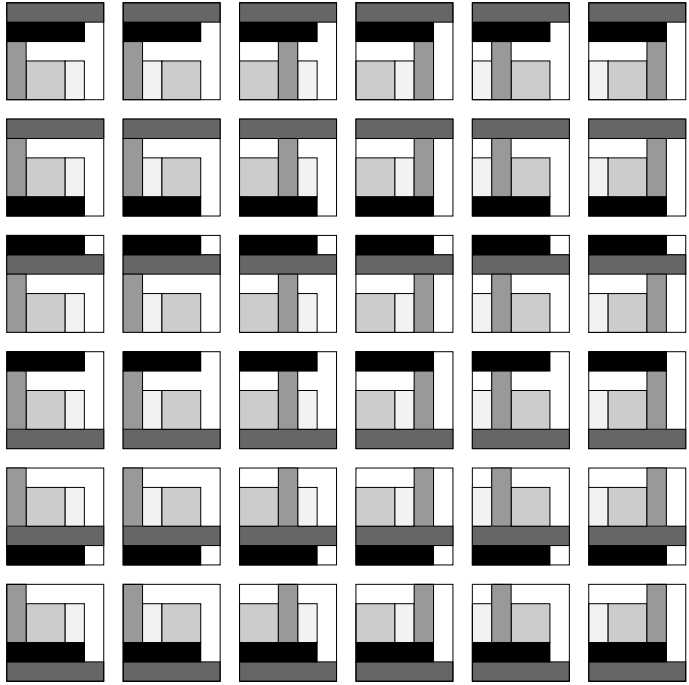}
\caption{\label{pmpic2} All feasible packings of the packing class $E$ from
Figure~\ref{fi:orient} (a).}
\end{center}
\end{figure}
\end{exam}

\begin{remark}
Independent from our work, an approach with some similarity to our 
packing classes can be found in an unpublished report
by \citeN{Jerrum}. In that report, two-dimensional
packings are described as pairs of complements of partial orders.
However, Jerrum does not observe that there is a purely combinatorial
characterization that allows it to map graphs to packings, as described in our
Lemma~\ref{pmsatz3}. The latter is the main result of this paper.
\end{remark}

\bigskip
We conclude this section by a graph-theoretic formulation
of a useful geometric property.

If a set of boxes $S$ has a total sum of $x_i$-widths exceeding
$k$ times the $x_i$-width of the container, then any feasible
packing must have an $x_i$--orthogonal cut that intersects at least
$k+1$ boxes. Using the terminology of packing classes, we
get an algorithmically useful formulation of this condition:
 
\begin{theorem} \label{pmclique1}
Let $E$ be a packing class for $(V,w)$, $i \in \{1,\dots,d\}$,
$G_{i} := (V,E_{i})$ and
$S \subseteq V$.
Then
$G_{i}[S]$ contains a clique of size
$\left\lceil\frac{\sum_{s\in S} w_{i}(s)}{W_i}\right\rceil$.
\end{theorem}
 
\proof
As an interval graph, $G_{i}$ is perfect.
Thus, the induced subgraph $G_{i}[S]$
has maximum clique size $\omega := \omega(G_{i}[S])$
equal to the chromatic number $\chi(G_{i}[S])$.
Hence, $S$  can be partitioned into
$\omega$ disjoint stable sets:
$S = \part_{k=1}^{\omega} S_{k}$.
All stable sets from $G_{i}[S]$ are also stable in $G_{i}$.
Then $P2$ implies for all $k \in \{1,\dots,\omega\}$ that
$w_{i}(S_{k}) \le 1$. Altogether, we get
$$
w_{i}(S) =
w_{i} \left(\bigcup_{k=1}^{\omega} S_{k}\right) =
\sum_{k=1}^{\omega} w_{d}(S_{k}) \le \omega.
$$
Because $\omega$ is integer, the claim follows.
\qed

A more elaborate discussion on bounds for higher-dimensional
packing and their applications can be found in \citeN{pack2}.
 
\section{Algorithmic Implications}
\label{sec:algo}

When trying to solve a packing problem to optimality, a crucial
step is deciding the orthogonal packing problem
(OPP) whether a particular set of boxes can be packed
in a feasible way. In a practical context, one will typically
use three different tools:

\medskip 
{\bf (1)} In order to see whether there is a fast positive answer 
we can try to find a packing by means of a heuristic. 

\medskip
{\bf (2)} A fast way to get a negative
answer is described in \citeN{pack2}, where
we discuss lower bounds for packing problems.

\bigskip
However, for hard instances, both of these approaches will fail
to provide an answer, meaning that a third tool has to be applied:

\medskip
{\bf (3)} If all else fails, we need to resort to an appropriate
enumeration method to solve the OPP.

\medskip 
In this section, we describe how packing classes can
be used to provide {\bf (3)} if both of the
easy approaches fail. (Because the OPP is NP-hard in the strong sense
(\citeN{GAJO79}), it is reasonable to use enumerative methods.) 

As we saw above,
the existence of a packing is equivalent
to the existence of a packing class. Furthermore, we have shown
that a feasible packing can be constructed from a packing class in
time that is linear in the number of edges.
This allows us to search for a packing class, instead of a packing.
As we describe in the following, the advantage of this approach
lies not only in exploiting the symmetries arising from
different transitive orientations of the same packing class,
but also in the fact that the structural properties of packing classes
give rise to very efficient rules for identifying irrelevant
portions of the search tree.
 
Some of the technical implementation
details of the enumeration scheme are rather involved;
they are described in \cite{pack3}. Here we give an 
overview over the underlying mathematical ideas.

\subsection{Building Packing Classes}
When trying to build a packing class $E_1,\ldots,E_d$ 
for a set of boxes, we use a branch-and-bound approach. 
The algorithm branches by fixing any edge
$\{b,c\} \in E_{i}$ or by excluding it via
$\{b,c\} \notin E_{i}$. The set of edges that are ``positively fixed''
 to be in some $E_i$ is called $E_{+}$,
while the set of ``negatively fixed'' edges that are 
excluded are called $E_{-}$.
Fixing an edge to be in $E_{+}$ is called
{\em augmenting $E_{+}$}, while
excluding it means {\em augmenting $E_{-}$}.
When referring to a particular $E_i$, the respective
sets of included and excluded edges are denoted by $E_{+, i}$ and $E_{-, i}$.

The branching has two possible objectives:
\begin{enumerate}
\item[{\bf (Y)}]
Augment $E_{+}$ until it is a packing class.
\item[{\bf (N)}]
Prove that no augmentation of
$E_{+}$ to a packing class is possible, as it would have to
use ``excluded '' edges from $E_{-}$.
\end{enumerate}
In the first case, our tree search has been successful.
In the second case, the search on the current subtree may be terminated,
because the search space is empty.
Otherwise, we have to continue branching
until one of the two objectives is reached.

In the following we show how graph-theoretic properties
can be exploited to help with these objectives.
 
\subsection{Excluded Induced Subgraphs} \label{induzsubsection}
Following the idea described in the previous paragraph, we
need three components for our enumeration scheme:

{\bf (1)} A test
``Is the current set of fixed edges $E_{+}$ a packing class?''

{\bf (2)} A sufficient criterion that
$E_{+}$ has no feasible augmentation.

{\bf (3)} A construction method for feasible augmentations.

All three of these components can be reduced to
identifying or avoiding particular induced subgraphs.

First of all, it is easy to determine all
edges that are excluded by condition $P3$.
By performing these augmentations of
$E_{-}$ immediately, we can guarantee that
$P3$ is satisfied. Thus we can assume that $P3$ is satisfied,
and will remain satisfied by further augmentations.
 
$P2$ explicitly excludes certain induced subgraphs:
$i$-infeasible stable sets, i.\,e., $i$-infeasible cliques in the complement
of each component graph.
 
In order to formulate $P1$ in terms of excluded induced
subgraphs, we use the following powerful
Propositions~\ref{ivchar1} and \ref{kompchar}
-- the reader is referred to
the book by \citeN{GOL80}. 
In this context, we use a couple of technical terms:

\begin{definition}[Chords, Cocomparability Graphs]\ \\
For a cycle $C := [b_{0}, \dots, b_{k-1}, b_{k}=b_{0}]$ of length $k$,
the edges $b_{i}b_{j}, i,j \in \{0,\dots,k-1\}$
with $(|i-j| \mbox{ mod } k) > 1$ are called {\em chords}; the chords
$b_{i}b_{j}, i,j \in \{0,\dots,k-1\}$ with $(|i-j| \mbox{ mod } k) = 2$
are called {\em $2$-chords} of $C$.
A cycle is ($2$-) {\em chordless}, iff it does not have any
($2$-) chords.

A graph is called {\em triangulated}, if it does not
contain any chordless cycle of length $k\geq 4$.

A graph $G=(V,E)$ is a {\em cocomparability graph}, if the complement
graph $G=(V,\overline{E})$ is a comparability graph.

In a graph $G=(V,E)$, $\{v_1,v_2,v_3\}\subseteq V$ 
is called an {\em asteroidal triple},
if any two of the vertices can be connected by a path that contains
no vertex adjacent to the third vertex.

\end{definition}

\begin{prop}[Gilmore and Hoffman 1964] \label{ivchar1}\ \\
A cocomparability graph is an interval graph, iff
it does not contain the chordless cycle $C_{4}$ of length 4
as an induced subgraph.
\end{prop}
 
\begin{prop}[Ghouil\`{a}-Houri 1962, Gilmore and Hoffman 1964] \label{kompchar}\ \\
A graph is a comparability graph, iff it does not contain a $2$-chordless
cycle of odd length.
\end{prop}

In the proof of Theorem~\ref{th:npc}, we will make use of yet another 
characterization of interval graphs.
 
\begin{prop}[Lekkerkerker and Boland 1962] \label{pr:asteroid}\ \\
A graph is an interval graph, iff it is a triangulated graph that does not
contain any asteroidal triple.
\end{prop}
 
%\begin{lemma}
%A graph is an interval graph, iff
%it does not contain any induced $C_{4}$ and its complement does not contain
%an odd $2$-chordless cycle.
%\end{lemma}
 
Thus, $E_{+}$ is a packing class, if for all
$i \in \{1,\dots,d\}$ the following holds
(recall that $P3$ is assumed to be satisfied):
\begin{enumerate}
\item $(V,E_{+, i})$ does not contain a $C_{4}$ as an induced
subgraph.
\item $(V,{E_{-,i}})$ does not contain
an odd $2$-chordless cycle.
\item $(V,{{E}_{-, i}})$ does not contain
an $i$-infeasible clique.
\end{enumerate}

For illustration, we provide the following example:

\begin{exam}
See Figure~\ref{fi:verboten}.
Consider $V=\{v_1,v_2,v_3,v_4\}$ and suppose that at some stage,
we have 
\[E_{+, i}=\{\{v_1,v_2\},\{v_3,v_4\},\{v_4,v_1\}\}\]
and 
\[E_{-, i}=\{\{v_2,v_4\},\{v_1,v_3\}\}.\]
This leaves the decision for the edge $\{v_2,v_3\}$.
Adding $\{v_2,v_3\}$ to $E_{+, i}$ would create an induced $C_4$
in $(V,E_{+, i})$; 
by Theorem~\ref{ivchar1}, this is not admissible, forcing
$\{v_2,v_3\}$ to be in $E_{-, i}$.

\begin{figure}[htbp]
\begin{center}
\leavevmode
\epsfxsize=.6\textwidth
\epsffile{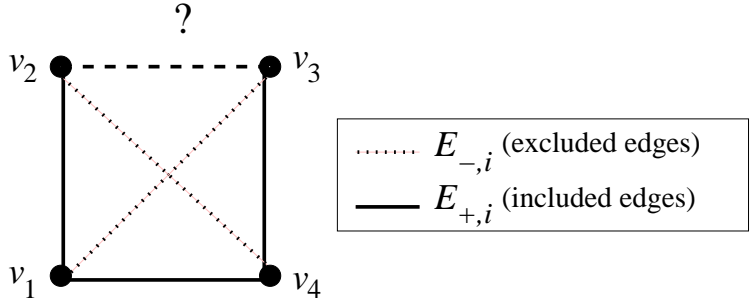}
\caption{\label{fi:verboten} Making use of Proposition~\ref{ivchar1}:
The edge $\{v_2,v_3\}$ must be added to $E_{-, i}$ to avoid an induced
$C_4$.}
\end{center}
\end{figure}

\end{exam}

In \cite{pack3}, we describe the technical details how
these properties can be implemented for achieving a fast algorithm.
The computational results described there show that the overall
code outperforms previous approaches by a wide margin.

\section{A Complexity Result}
\label{se:npc}
Tackling the OPP by building a packing class resolves
some of the difficulties, but of course there is little hope
of solving the problem in polynomial time by any method.
We conclude this paper by pointing out the exact nature of
the remaining difficulty.

Because properties $P1$ and $P3$ appear easy to deal with,
it is natural to suspect that this difficulty
lies in property $P2$. Thus, it seems reasonable to hope for
an efficient method that preserves properties $P1$ and $P3$
when augmenting a ``partial packing class'':
Given a $d$-tuple of edge sets that satisfies $P3$, can it be
augmented such that properties $P1$ and $P3$ are satisfied?
(Note that $P2$ would be inherited to all augmentations, because 
any stable set after an augmentation must have been 
a stable set before.)
A polynomial algorithm for this task would reduce the computational
difficulty of solving the OPP by means of enumeration.
 
Formally, we have the following problem:
 
\begin{definition}[Disjoint $d$-Interval Graph Completion Problem (DIGCP-$d$] \ \\
\indent
\begin{tabular}{ll}
GIVEN: & graphs ${\cal G}_{1} = (V,{\cal E}_{+,1}), \dots, {\cal G}_{d} = 
(V,{\cal E}_{+,d})$
with
$\bigcap_{i=1}^{d} {\cal E}_{+,i} = \emptyset$.\\
%\end{tabular}
 %
%\begin{tabular}{ll}
QUESTION: & For $i \in \{1,\dots,d\}$, is there
a superset $E_{i}$ for each ${\cal E}_{+,i}$, such that\\
& $\bigcap_{i=1}^{d} E_{i} = \emptyset$, and the
graphs
$G_{1}=(V,E_{1}), \dots, G_{d}=(V,E_{d})$ are interval \\
& graphs?
\end{tabular}
\end{definition}
 
%Sandwichtheorem \citeN{GOSH93}
 
In the dissertation of the second author~\cite{Sch97}
it is shown that the existence of a polynomial
algorithm for this problem is highly unlikely:
The OPP remains NP-hard, even if we have a $d$-tuple
of edge sets that satisfies $P2$ and $P3$.
This indicates that it is indeed $P1$ that makes life difficult.
 
\begin{theorem}
\label{th:npc}
For $d \ge 2$, the Disjoint $d$-Interval Graph Completion Problem
is NP-complete.
\end{theorem}

\proof
Recognizing interval graphs can be done in linear time~\cite{KOMO89},
so the problem is in NP.

For showing that the problem is NP-hard, we give a reduction of 
3SAT. In the following, all notation is consistent with \cite{GAJO79}.

\subsubsection*{Constructing a DIGCP-2 instance from a 3SAT instance:}

Consider a Boolean expression $\Phi$ in conjunctive normal form 
with variables $u_{1}, \dots, u_{n}$ and clauses
$c_{1}, \dots, c_{m}$.
The following DIGCP-2 instance can be constructed in polynomial time:

\begin{figure}[htb]
%\begin{minipage}[t]{8cm}
\begin{center}
\leavevmode
\epsfxsize=0.5\textwidth
\epsffile{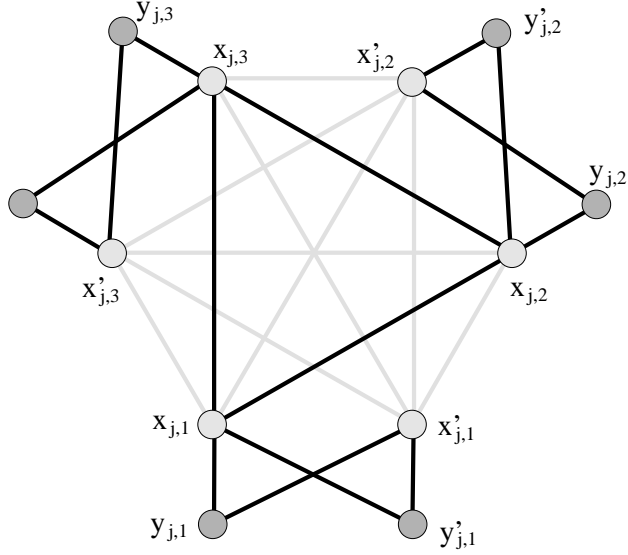}
\caption{The edges used for representing clauses.}
\label{sat3pic1}
\end{center}
\end{figure}
\begin{figure}[htb]
\begin{center}
\leavevmode
\epsfxsize=0.5\textwidth
\epsffile{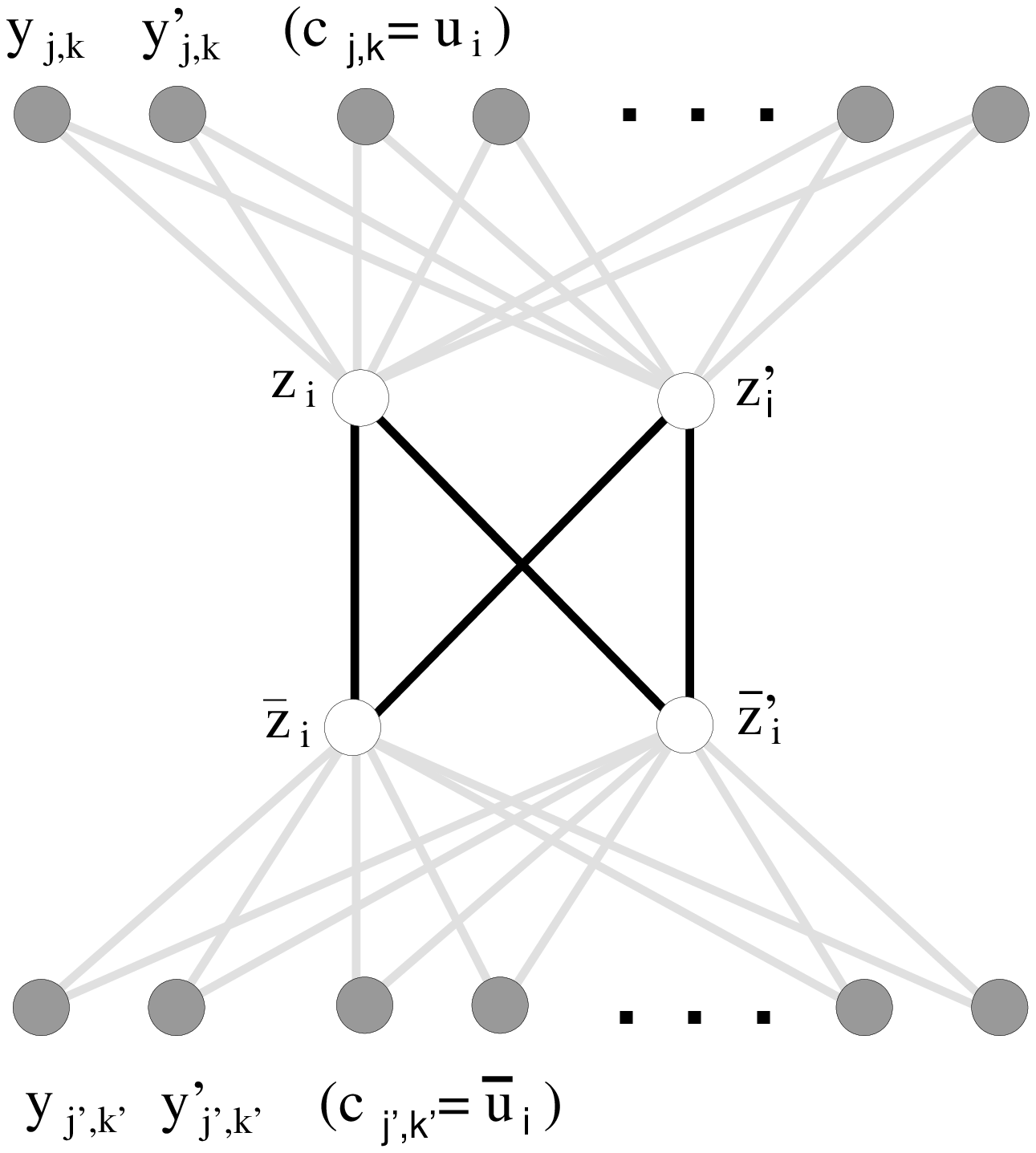}
\caption{The edges used for representing variables.}
\label{sat3pic2}
\end{center}
\end{figure}

\begin{enumerate}
\item For each clause $c_{j}$ define six
{\em interior} vertices
$x_{j,k},x_{j,k}', \, k \in \{ 1,\dots,3\}$ and
six {\em exterior} vertices
$y_{j,k},y_{j,k}', \, k \in \{1,\dots,3,\}$.
Connect these vertices according to
Figure~\ref{sat3pic1}.
The black edges form the set $C_{j}^{(1)}$, and the gray edges
form the set $C_{j}^{(2)}$. Thus we have
\begin{eqnarray*}
C_{j}^{(1)} & = &
\{ \{x_{j,1}x_{j,2}\},\{ x_{j,1}x_{j,3}\}, \{x_{j,2}x_{j,3}\} \}
\cup
\bigcup_{k=1}^{3}
\{
x_{j,k}y_{j,k},
x_{j,k}y_{j,k}',
x_{j,k}'y_{j,k},
x_{j,k}'y_{j,k}'
\},
\\
C_{j}^{(2)} & = &
\{ x_{j,1}'x_{j,2}', x_{j,1}'x_{j,3}', x_{j,2}'x_{j,3}' \}
\cup
\bigcup_{k=1}^{3}
\bigcup_{\scriptstyle l=1 \atop \scriptstyle l \ne k}^{3}
\{
x_{j,k}x_{j,l}'
\}.
\end{eqnarray*}

\item For each 
variable $u_{i}$ define the four vertices 
$z_{i},z_{i}', \bar{z}_{i},\bar{z}_{i}'$.
If $u_{i}$
is the $k$th literal in clause $c_{j}$, then
$z_{i},z_{i}'$ are connected with the exterior vertices 
$y_{j,k}$ and $y'_{j,k}$ of
$c_{j}$.
If $\bar{u}_{i}$
is the $k$th literal of clause $c_{j}$,
$\bar{z}_{i},\bar{z}_{i}'$ are connected with
$y_{j,k}$ and $y'_{j,k}$.
Thus, we get the two edge sets
$$
U_{i}^{(2)}  :=
\{
z_{i}
y_{j,k}
,
z_{i}'
y_{j,k}
,
z_{i}
y_{j,k}'
,
z_{i}'
y_{j,k}'
| \quad \mbox{ $u_{i}$ is the $k$-th literal of $c_{j}$}
\}
$$
and
$$
\bar{U}_{i}^{(2)}  =
\{
\bar{z}_{i}
y_{j,k}
,
\bar{z}_{i}'
y_{j,k}
,
\bar{z}_{i}
y_{j,k}'
,
\bar{z}_{i}'
y_{j,k}' \quad
| \quad \mbox{ $\bar{u}_{i}$ is the $k$-th literal of $c_{j}$}
\}
$$

These edges are shown gray in Figure~\ref{sat3pic2}.
The black edges form the set
$$
U_{i}^{(1)}  =
\{
z_{i}\bar{z}_{i},
z_{i}\bar{z}_{i}',
z_{i}'\bar{z}_{i},
z_{i}'\bar{z}_{i}'
\}.
$$
\end{enumerate}

Now the
DIGCP-$2$ instance $I$ is given by
\begin{eqnarray*}
V & := &
\bigcup_{j=1}^{m}
\bigcup_{k=1}^{3}
\left\{
x_{j,k},x_{j,k}',
y_{j,k},y_{j,k}' \right\}
\, \cup \,
\bigcup_{i=1}^{n} \left\{
z_{i},z_{i}',
\bar{z}_{i},\bar{z}_{i}'  \right\} \\
{\cal E}_{+,1} & := &
\bigcup_{j=1}^{m} C_{j}^{(1)} \, \cup \,
\bigcup_{i=1}^{n} U_{i}^{(1)}  \\
{\cal E}_{+,2} & := &
\bigcup_{j=1}^{m} C_{j}^{(2)} \, \cup \,
\bigcup_{i=1}^{n} U_{i}^{(2)} \, \cup \,
\bigcup_{i=1}^{n} \bar{U}_{i}^{(2)}.
\end{eqnarray*}

\subsubsection*{Constructing a solution for 3SAT from a solution of DIGCP-2:}
Let now 
$(E_{1}, E_{2})$ be a solution of $I$.
For $i \in \{1,\dots,n\}$ let
$$
u_{i} := \left\{
\begin{array}{ll}
\mbox{\tt TRUE}, &
z_{i}z_{i}' \notin E_{1} \\
\mbox{\tt FALSE}, &
z_{i}z_{i}' \in E_{1}
\end{array}
\right.
$$

In order to show that this assignment of variables satisfies
$\Phi$, we will repeatedly use Proposition~\ref{pr:asteroid}.
In particular, the proposition implies that an interval graph
cannot contain an asteroidal triple or a $C_4$ as an induced subgraph.

If $u_{i} = \mbox{\tt TRUE}$, hence
$z_{i},z_{i}' \notin E_{1}$, then
$\bar{z}_{i}\bar{z}_{i}' \in E_{1}$ must hold;
Otherwise the edge set
$
\{z_{i},
\bar{z}_{i},
z_{i}',
\bar{z}_{i}'\}$ would induce a $C_{4}$ in $(V,E_{1})$.

Let $c_{j}$ be a clause in $\Phi$.
In Figure~\ref{sat3pic1} consider the subgraph induced in ${\cal E}_{+,2}$
by the interior vertices of $c_{j}$ that has the edge set 
$C_{j}^{(2)}$. We note the following.

\begin{enumerate}
\item No vertex of the path $(x_{j,1}, x_{j,3}', x_{j,2})$ is adjacent to
$x_{j,3}$ by an edge in $C_{j}^{(2)}$
\item No vertex of the path $(x_{j,1}, x_{j,2}', x_{j,3})$ is adjacent to
$x_{j,2}$ by an edge in $C_{j}^{(2)}$
\item No vertex of the path $(x_{j,2}, x_{j,1}', x_{j,1})$ is adjacent to
$x_{j,1}$ by an edge in $C_{j}^{(2)}$
\end{enumerate}
Thus, the vertices $x_{j,1}$, $x_{j,2}$, $x_{j,3}$  form an asteroidal triple,
unless one or several edges prevent this by introducing adjacencies.
More precisely, $E_{2}$ must contain at least one of the edges
$x_{j,1}x_{j,1}'$,
$x_{j,2}x_{j,2}'$,
$x_{j,3}x_{j,3}'$  in addition to
$C_{j}^{(2)}$, because all other edges that could destroy the
asteroidal triple are in ${\cal E}_{+,1}$.

Without loss of generality let $x_{j,1}x_{j,1}' \in E_{2}$
and hence $x_{j,1}x_{j,1}' \notin E_{1}$.
In order to prevent the four vertices
$\{ x_{j,1},y_{j,1},x_{j,1}',y_{j,1}' \}$ from inducing a $C_{4}$ in
$(V,E_{1})$, 
$y_{j,1}y_{j,1}' \in E_{1}$ must hold.
Using the same argument on $E_{2}$,
$y_{j,1}y_{j,1}' \notin E_{2}$ implies that the edge 
$z_{i}z_{i}'$ (or $\bar{z}_{i}\bar{z}_{i}'$, resp.)
belonging to the first literal $u_{i}$ 
(or $\bar{u}_{i}$, resp.) must be in
$E_{2}$, hence not in $E_{1}$.
Corresponding to our chosen truth assignment,
we get $u_{i} = \mbox{\tt TRUE}$ (or $u_{i} = \mbox{\tt FALSE}$, resp.) 
Therefore, clause
$c_{j}$ is satisfied for any $j$, and hence $\Phi$.

\subsubsection*{Constructing a solution for DIGCP-2 from a solution of 3SAT.}
Consider a truth assignment of the $u_{i}$ that satisfies $\Phi$.
Without loss of generality, let the first literal in each clause
be {\tt TRUE}, and let $i \in \{1, \dots, n\}$.

If $u_{i} = {\tt TRUE}$, let 
\begin{eqnarray*}
\hat{U_{i}}^{(1)} & := &
U_{i}^{(1)}  \cup
\{ \bar{z}_{i}\bar{z}_{i}' \}, \\
\hat{U_{i}}^{(2)} & := &
U_{i}^{(2)}  \cup
\{ z_{i}z_{i}' \}, \\
\hat{\overline{U_{i}}}^{(2)} & := &
\bar{U_{i}}^{(2)}
\cup \clique{A} \mbox{ with }
A := \{ y_{j,k}, y_{j,k}' | \mbox{ $\bar{u}_{i}$ is the $k$-th literal of $c_{j}$} \}.
\end{eqnarray*}

If $u_{i} = {\tt FALSE}$, let 
\begin{eqnarray*}
\hat{U_{i}}^{(1)} & := &
U_{i}^{(1)}  \cup
\{ z_{i}z_{i}' \}, \\
\hat{U_{i}}^{(2)} & := &
U_{i}^{(2)}  \cup
\{ \bar{z}_{i}\bar{z}_{i}' \}, \\
\hat{\overline{U_{i}}}^{(2)} & := &
\bar{U_{i}}^{(2)}
\cup \clique{B}
\mbox{ with }
B := \{ y_{j,k}, y_{j,k}' | \mbox{ $u_{i}$ is the $k$-th literal of $c_{j}$} \}.
\end{eqnarray*}

It is easy to see that the corresponding subgraphs are interval graphs:

For $j \in \{1, \dots, m\}$, let
\begin{eqnarray*}
\hat{C_{j}}^{(1)} & := &
C_{j}^{(1)} \cup
\{ y_{j,1}y_{j,1}',
x_{j,2}x_{j,2}',
x_{j,3}x_{j,3}' \}
\, \cup \,
\{ y_{j,1}, y_{j,1}' \} \times
\{ x_{j,2},
x_{j,2}',
y_{j,2},
y_{j,2}' \} \\
\hat{C_{j}}^{(2)} & := &
C_{j}^{(2)} \cup
\{ (x_{j,1}x_{j,1}') \}
\end{eqnarray*}

\begin{figure}[htb]
\begin{center}
\leavevmode
\epsfxsize=0.5\textwidth
\epsffile{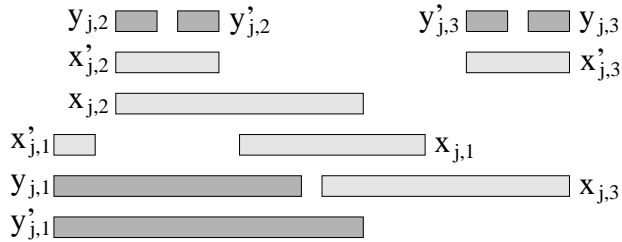}
\caption{Representing $(V, \hat{C_{j}}^{(1)})$ by intervals.}
\label{irep1}
\end{center}
\end{figure}

\begin{figure}[htb]
\begin{center}
\leavevmode
\epsfxsize=0.5\textwidth
\epsffile{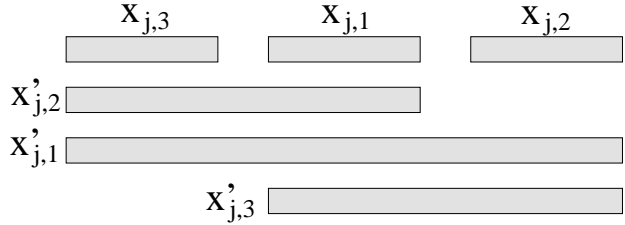}
\caption{Representing $(V, \hat{C_{j}}^{(2)})$ by intervals.}
\label{irep2}
\end{center}
\end{figure}

The representations shown in
Figures~\ref{irep1} and \ref{irep2} show
that $(V, \hat{C_{j}}^{(1)})$ and
$(V, \hat{C_{j}}^{(2)})$ are also interval graphs.


Let now
\begin{eqnarray*}
E_{1} & := &
\bigcup_{i=1}^{n} \hat{U_{i}}^{(1)} \cup \bigcup_{j=1}^{m} \hat{C_{j}}^{(1)}, \\
E_{2} & := & \bigcup_{i=1}^{n} \hat{U_{i}}^{(2)} \cup
\bigcup_{i=1}^{n} \hat{\overline{U_{i}}}^{(2)} \cup
\bigcup_{j=1}^{m} \hat{C_{j}}^{(1)}.
\end{eqnarray*}

$(V,E_{1})$ and
$(V,E_{2})$ are interval graphs, because all their connected components are.
${\cal E}_{+,1} \subseteq E_{1}$ and
${\cal E}_{+,2} \subseteq E_{2}$ follows immediately from the construction.

For $j \in \{1,\dots,m\}$, let $u_{i}$ be the first literal from clause
$c_{j}$.
By assumption, this literal has the value
{\tt TRUE}, implying that the edge
$y_{j,1}y_{j,1}'$ does not belong to 
$\hat{U}_{i}$ or
$\hat{\overline{U}}_{i}$, resp. 
Therefore, $\hat{C}_{i}$ is disjoint from these edge sets.
This excludes the only possible overlap of 
$E_{1}$ and $E_{2}$,
implying
$E_{1} \cap E_{2} = \emptyset$.
Therefore, $(E_{1}, E_{2})$ is a solution of the DIGCP-2 instance $I$.
\qed

\section{Conclusions}
\label{sec:conc}
In this paper, we have introduced a new way of characterizing
higher-dimensional packings of a set of boxes into a container.
Using graph-theoretic structures, this characterization has a number
of nice algorithmic properties. We describe in
\cite{pack3} how these properties can be used
for implementing an exact algorithm that is able to solve
two- and three-dimensional packing problems of relevant size.
Our implementation makes extensive use of new classes of
lower bounds for higher-dimensional packing problem; the mathematical
background is described in \citeN{pack2}.
The overall algorithm appears to have convincing performance; 
this should make the mathematical structures
and properties described in this paper interesting to a wide audience.

Our graph-theoretic characterization has also turned out to be useful
in practical applications in which there are additional 
spatial constraints in some of the dimensions. See 
\citeN{Teich} for an application arising from dynamic reconfiguration
of hardware. Making further use of order-theoretic structures, we
have been able to extend this approach to handle order constraints.
Details are described in \citeN{pack4}.

\section*{Acknowledgments}
We thank Mark Jerrum for pointing out his previous paper \cite{Jerrum},
and for encouraging comments on the extension of our work.
We also thank two anonymous referees for comments that helped to improve
the presentation of this paper.

A previous extended abstract version summarizing the results
of this paper in the version \cite{pack1}, and references 
\cite{pack2} and \cite{pack3} appears in
{\em Algorithms -- ESA'97}~\cite{esa}. This work originated from the second
author's doctoral thesis~\cite{Sch97},
supported by the German Federal Ministry of Education,
Science, Research and Technology (BMBF, F\"orderkennzeichen 01~IR~411~C7).

\bibliographystyle{chicago}
\bibliography{packing}

\end{document}